\documentclass[5p,times,a4paper,fleqn]{cas-dc}

\usepackage[numbers]{natbib}

\usepackage{todonotes}
\usepackage[linesnumbered,ruled,vlined]{algorithm2e}
\usepackage{ulem}
\usepackage{multirow}
\usepackage{float}
\restylefloat{table}

\def\tsc#1{\csdef{#1}{\textsc{\lowercase{#1}}\xspace}}
\tsc{WGM}
\tsc{QE}
\tsc{EP}
\tsc{PMS}
\tsc{BEC}
\tsc{DE}

\begin{document}
\let\WriteBookmarks\relax
\def\floatpagepagefraction{1}
\def\textpagefraction{.001}

\shorttitle{Extending DD-$\alpha$AMG on heterogeneous machines}

\shortauthors{Ramirez-Hidalgo, He, Zhang}

\title[mode = title]{Extending DD-$\alpha$AMG on heterogeneous machines}

%

\author[1]{Gustavo Ramirez-Hidalgo}[]
\ead{g.ramirez.hidalgo@fz-juelich.de}

\affiliation[1]{organization={J{\"u}lich Supercomputing Centre, Forschungszentrum J{\"u}lich GmbH},
    addressline={Wilhelm-Johnen-Straße},
    city={Jülich},
    postcode={52428},
    country={Germany}}

\author[2]{Lianhua He}[]
\ead{helh@sccas.cn}
\affiliation[2]{organization={Computer Network Information Center, Chinese Academy of Sciences},
    addressline={CAS Informatization Plaza No.2 Dong Sheng Nan Lu, Haidian District},
    city={Beijing},
    country={China}}

\author[2]{Ke-Long Zhang}
\ead{klzhang@cnic.cn}


\begin{abstract}
Multigrid solvers are the standard in modern scientific computing simulations. Domain Decomposition Aggregation-Based Algebraic Multigrid, also known as the DD-$\alpha$AMG solver, is a successful realization of an algebraic multigrid solver for lattice quantum chromodynamics. Its CPU implementation has made it possible to construct, for some particular discretizations, simulations otherwise computationally unfeasible, and furthermore it has motivated the development and improvement of other algebraic multigrid solvers in the area. From an existing version of DD-$\alpha$AMG already partially ported via \texttt{CUDA} to run some finest-level operations of the multigrid solver on Nvidia GPUs, we translate the \texttt{CUDA} code here by using \texttt{HIP} to run on the ORISE supercomputer. We moreover extend the smoothers available in DD-$\alpha$AMG, paying particular attention to Richardson smoothing, which in our numerical experiments has led to a multigrid solver faster than smoothing with GCR and only 10\% slower compared to SAP smoothing. Then we port the odd-even-preconditioned versions of GMRES and Richardson via \texttt{CUDA}. Finally, we extend some computationally intensive coarse-grid operations via advanced vectorization.
\end{abstract}


\maketitle

\section{Introduction}

The strong force, one of the four fundamental forces in nature, is modelled by the theory of quantum chromodynamics (QCD) \cite{peskin2018introduction}. Due to the model itself, QCD cannot be solved analytically in some energetic regimes, and in order to obtain predictions under such conditions one has to employ numerical methods via what is known as Lattice QCD (LQCD) \cite{gattringer2009quantum}. LQCD simulations are amongst the most demanding applications in scientific computing, in terms of High Performance Computing (HPC) usage. Furthermore, one of the most frequent and expensive ``fundamental" operations within LQCD simulations is the solution of linear systems of the form $Dx = b$, where $D$ is the Dirac operator. The form of $D$ varies depending on the discretization procedure, leading to different formulations of LQCD, such as Wilson, twisted mass, staggered, and others \cite{gattringer2009quantum,wilson1974confinement,frezzotti2000local,kogut1975hamiltonian}.

Multigrid methods are the state of the art when solving linear systems in LQCD. In particular, using an algebraic multigrid (AMG) method \cite{trottenberg2000multigrid} rather than a geometric one is a necessity for this problem. The current status of AMG solvers in LQCD comes from twenty years of continuous algorithmic and computational developments \cite{Babich2010,Brannick2008,
Brower2020,Frommer2014,osborn2010multigrid}.

One realization of an AMG solver in the context of LQCD is DD-$\alpha$AMG \cite{rottmann2016adaptive,Frommer2014}.
This solver, originally developed as a pure CPU code in \texttt{C}, parallelized via \texttt{MPI} and \texttt{OpenMP} and vectorized with \texttt{SSE}, has recently seen some of its most expensive sections ported to run on GPUs via \texttt{CUDA}
\cite{RamirezHidalgo_PhDThesis,arxivAcceleratingLattice}. We start here from this hybrid CPU+GPU version and extend it by using \texttt{HIP} to run on the ORISE supercomputer, then we extend the \texttt{C} and \texttt{CUDA} codes both algorithmically and computationally, and finally upgrade some coarse-grid kernels to support \texttt{AVX2} and \texttt{AVX-512} instead of \texttt{SSE}.

This paper is distributed as follows. Sect.~\ref{sect:amg_and_orise} sets the background by describing the ORISE supercomputer, AMG and DD-$\alpha$AMG, followed by a description and comparison of the iterative methods relevant for the extensions of DD-$\alpha$AMG done in this work, and finally we describe and compare \texttt{CUDA} and \texttt{HIP}. Then, sect.~\ref{sect:num_exps} presents numerical experiments where we discuss the extensions to DD-$\alpha$AMG realized here, with runs done on ORISE, and we end this section with a discussion on the impact of the CPU improvements via \texttt{AVX}. Finally, we present conclusions and current and future work in sect.~\ref{sect:conclusions_and_future_work}.

\section{Multigrid on ORISE via \texttt{HIP}}\label{sect:amg_and_orise}

In this work we, among other things, translate the current CPU+GPU version of DD-$\alpha$AMG \cite{wuppertalGPUDDalphaAMG} via the Heterogeneous-Compute Interface for Portability (\texttt{HIP}) \cite{HIPdocum88:online,HIP_GitHub}. The numerical experiments presented in subsect.s \ref{subsect:smoothers_comp_cpu_only} and \ref{subsect:ddalphaamg_on_orise_results} run on the ORISE supercomputer, hence we start this section by describing the architecture in ORISE. This is followed by a brief introduction to AMG and to the AMG solver DD-$\alpha$AMG, which is being extended here. We then do a comparative description of two iterative methods available in DD-$\alpha$AMG as smoothers: the Generalized Minimal Residual (GMRES) method and Schwarz Alternating Producedure (SAP), discussing in particular their pros and cons in terms of smoothing and of running on CPUs and GPUs; we extend the discussion (and DD-$\alpha$AMG itself) by adding two more iterative methods to the stack of smoothers: Generalized Conjugate Residual (GCR) and modified Richardson. We finalize this section by briefly describing \texttt{CUDA} and \texttt{HIP}, emphasizing their differences and similarities.

\subsection{The ORISE Supercomputer}

This machine consists of CPU nodes equipped with GPUs. On the CPU side, each node is 4-way 8-core, for a total of 32 cores, with an x86 instruction set and 128 GB of RAM. It supports the POSIX standard and common parallel programming models such as MPI and OpenMP. Four GPUs are attached to each node. A single GPU is composed of 64 streaming multiprocessors and 16 GB of HBM2 RAM.  Each node is connected to its GPU accelerators via 32 PCIe buses. Each accelerator accesses its corresponding CPU through direct memory access (DMA) with 16 GB/s bandwidth, and the nodes are linked by a high-speed network with 25 GB/s bandwidth.
ORISE is capable of using \texttt{HIP} to run \texttt{CUDA}-based applications on its GPUs.

Direct GPU-to-GPU communication is not possible in ORISE, hence we resort in our implementations to the more expensive procedure of doing GPU-to-CPU followed by CPU-to-GPU in order to be able to communicate data from a GPU to another.

\subsection{Algebraic Multigrid}\label{subsect:amg}

When solving linear systems $Dx = b$ in general, traditional iterative methods such as Jacobi, Gauss-Seidel, GMRES, etc.~\cite{saad2003iterative}, although different from each other in many aspects, they all share a common drawback: as the condition number of the matrix $D$ i.e.\ $\kappa(D)$, increases, the iteration count for them to reach a certain relative tolerance goes up, sometimes very fast. On the other hand, as parameters in scientific computing simulations are moved closer to their continuum counterparts, the condition number of some matrices appearing in linear systems, to be solved in those simulations, grows rapidly.

Multigrid methods \cite{trottenberg2000multigrid} are born out of the desire to have a solver whose convergence is independent of $\kappa(D)$. To achieve this independence, \textit{smoothers} and \textit{coarse grids} are combined. Smoothers consist of a few iterations of traditional methods (e.g.\ Gauss-Seidel, GMRES). With just a few iterations, such methods quickly remove error components associated to large eigenvalues\footnote{For the remainder of this subsection, we denote by $e_{low}$ that part of the error approximately connected to small eigenvalues (in magnitude) of $D$ and $e_{high}$ that connected to large eigenvalues.} of $D$, but they stagnate at some point as components of the error corresponding to small eigenvalues of $D$ are not easy to deal with for those algorithms. That is when multigrid methods become beneficial or even necessary. After a few iterations of a traditional method, a coarse grid is used and information related to the error is transferred from the fine to the coarse grid. An approximation to the error is computed on the coarse grid, then transferred back to the fine grid and the current approximate corrected. Let us depict multigrid as a step-by-step procedure now:
\begin{enumerate}
    \item Set $i \leftarrow 0$ and an initial guess $x^{(0)}$.
    \item Set $i \leftarrow i+1$. Set $x^{(i)} \leftarrow x^{(i-1)}$. Update the \textit{residual} $r^{(i)} \leftarrow b - D x^{(i)}$, check the stopping condition and stop if necessary. Based on the relation between the error and the residual $D e^{(i)} = r^{(i)}$, perform $\nu$ iterations of pre-smoothing $M^{(\nu)}$, obtaining with this $\widehat{e}^{(i)} \leftarrow M^{(\nu)} r^{(i)}$. Then correct the current iterate i.e.\ $x^{(i)} \leftarrow x^{(i)} + \widehat{e}^{(i)}$. The smoother $M^{(\nu)}$, being effective in approximately suppressing $e_{high}$, renders the new error $e^{(i)}$ less rich in large eigenmodes of $D$.
    \item Update the residual $r^{(i)} \leftarrow b - D x^{(i)}$. With linear transformations $P$ and $R$ transferring quantities to and from the coarse grid, respectively, a \textit{coarse grid correction} of the form $x^{(i)} \leftarrow x^{(i)} + \widehat{e}^{(i)}$ with $\widehat{e}^{(i)} \leftarrow P D_{c}^{-1} R r^{(i)}$ and $D_{c} = R D P$, which is effective in approximately removing $e_{low}$, leads to a new error $e^{(i)}$ with many of its error components related to small eigenmodes of $D$ further suppressed.
    \item Update the residual $r^{(i)} \leftarrow b - D x^{(i)}$. Post-smoothing is applied in the same way as pre-smoothing, i.e.\ we get $x^{(i)} \leftarrow x^{(i)} + \widehat{e}^{(i)}$ with $\widehat{e}^{(i)} \leftarrow M^{(\nu)} r^{(i)}$. Check stopping criterion, stop if finished otherwise go to step 2.
\end{enumerate}

The two-level method listed above can be further used in a recursive manner, as the matrix $D_{c}$ in the operation $D_{c}^{-1} R r^{(i)}$ might be too large to invert with direct methods or too ill-conditioned to use traditional methods e.g.\ restarted GMRES, hence we might want to use another two-level method, recursively, to solve the system $D_{c} x_{c} = R r^{(i)}$, rendering the method a multilevel one.

Whether we are working with \textit{geometric} \cite{briggs2000multigrid,wesseling2001geometric} or \textit{algebraic} \cite{stuben2001introduction,falgout2006introduction} multigrid depends on what information we use to build $P$ and $R$. In geometric multigrid the structure of the grid is the sole ingredient needed, e.g.\ one can design $P$ such that it interpolates (i.e.\ averages) values on lattice sites when going from the coarse to the fine grid, and then one can choose for example $R = P^{H}$. This choice of $R = P^{H}$ is known as Galerkin construction \cite{briggs2000multigrid}.

In an algebraic multigrid method, using data from the matrix $D$ to build $P$ and $R$ is fundamental. Let us briefly go into this in more detail. Let us assume that the span of the columns of $P$ approximates the subspace spanned by many small eigenmodes of $D$, i.e.\ $\texttt{range}(P) \approx \texttt{range}(V)$, where the columns of $V$ contain many of the smallest eigenvectors of $D$. Then, we can write $e_{low} \approx Pu$ for some vector $u$. The coarse grid correction, i.e.\ $x^{(i)} \leftarrow x^{(i)} + P D_{c}^{-1} R r^{(i)}$, can be equivalently written in terms of the error update as $e^{(i)} \leftarrow (I - P D_{c}^{-1} R D) e^{(i)}$, which leads then in turn to $e_{low}^{(i)} \leftarrow (I - P D_{c}^{-1} R D) e_{low}^{(i)} \approx Pu - P D_{c}^{-1} R D Pu = 0$. Hence we see that having $\texttt{range}(P)$ rich in small eigenmodes of $D$ gives us a good coarse grid correction. In the next subsection a realization of an AMG solver for LQCD is discussed.

\subsection{Domain Decomposition Aggregation-Based Algebraic Multigrid}\label{subsect:ddalphaamg}

Domain decomposition aggregation-based algebraic multigrid (DD-$\alpha$AMG) \cite{rottmann2016adaptive,Frommer2014} is both an algorithmic framework and a code \cite{wuppertalDDalphaAMG} for solving linear systems in LQCD. It targets the Wilson-Dirac discretization with a clover improvement.

In the case of the Dirac operator of LQCD, an algebraic construction of $P$ from $D$ is needed. DD-$\alpha$AMG implements this via an \textit{aggregation-based} AMG. This aggregation-based construction relies on the concept of \textit{local coherence} \cite{luscher2007local}, which states that many low modes of $D$ can be approximately obtained from just a few low modes by looking at the local behaviour of those few modes. This construction allows us to have many approximate low modes of $D$ as columns of $P$, with a sparse structure in $P$ leading to a coarse-grid Dirac operator $D_{c}$ resembling $D$ in its nearest-neighbor structure. At the fine grid, the degrees of freedom (d.o.f.) per lattice site is twelve hence $D$ has nine $12 \times 12$ blocks every twelve rows, and for $D_{c}$ this structure is the same except we have now nine $2N_{v} \times 2N_{v}$ blocks every $2N_{v}$ rows. The quantity $N_{v}$ is called the number of \textit{test vectors}, which are the approximate low modes that we use to construct $P$; a typical value is $N_{v} = 24$. In a multilevel setting (i.e.\ more than two levels) the next coarse grid is built recursively by approximating low modes of $D_{c}$.

The DD-$\alpha$AMG solver has a \textit{setup} and a \textit{solve} phase. During the setup phase it builds the \textit{multigrid hierarchy}. Let us define this multigrid hierarchy as the set of operators $D_{\ell}$ with $\ell = 1, ..., L$, and $P_{\ell}$ and $R_{\ell} = P_{\ell}^{H}$ with $\ell = 1, ..., L-1$, $L$ being the total number of levels. The solve phase then makes use of this hierarchy of operators to solve one or more linear systems i.e.\ for one or more right hand sides in a sequence.

During the solve phase, DD-$\alpha$AMG makes use of, at each level but its coarsest, flexible GMRES \cite{saad2003iterative} (FGMRES) preconditioned by a two-level AMG. The FGMRES at the fine grid is solved up to the desired relative residual tolerance e.g.\ $10^{-12}$. If two levels are used, the coarse-grid linear system is solved via GMRES up to a relative residual tolerance of $10^{-1}$; if more than two levels are under use, then FGMRES is used instead of GMRES, still up to $10^{-1}$, but now preconditioned by another two-level method, leading to a three-level solver. This cycling strategy i.e.\ using FMGRES to wrap every level of the multigrid solver is called a K-cycle \cite{notay2008recursive}.

In DD-$\alpha$AMG, (F)GMRES at coarser levels is usually solved up to $10^{-1}$. For the smoother one can choose one of either GMRES or SAP, and we will discuss further on this in subsect.~\ref{subsect:smoothers_comp}. If the parameters of the multigrid solver are chosen appropriately, as the conditioning of $D$ becomes larger the number of iterations for the finest-level FGMRES to reach the desired tolerance will remain basically constant. With increasing $\kappa(D)$ come more ill-conditioned coarser levels, which impacts in particular the coarsest level i.e.\ as $\kappa(D)$ increases we will see increasing execution times for the solver to converge, mostly due to an increasing iteration count at the coarsest level to reach the desired $10^{-1}$. One can equip the coarsest level with preconditioning and deflation to partially restore insensitivity to $\kappa(D)$ in the total execution time of the solver, as discussed in \cite{espinoza2023coarsest}.

\subsection{The smoothers}\label{subsect:smoothers_comp}

In the original version of DD-$\alpha$AMG \cite{wuppertalDDalphaAMG}, a CPU-only code, two smoothers were made available: GMRES and SAP. In the more recent CPU+GPU version \cite{wuppertalGPUDDalphaAMG}, SAP at the finest level was ported via \texttt{CUDA} to run on GPUs, while the rest of the code was kept executing on CPUs. In this work, we have extended the CPU version to support GCR and modified Richardson as smoothers. Furthermore, we have implemented \texttt{CUDA} code so that the solver is now able to run with SAP, GMRES or Richardson offloaded to GPUs.

Numerical tests comparing the different smoothers are presented in subsect.~\ref{subsect:smoothers_comp_cpu_only}. We now describe, discuss and compare the four smoothers considered in those tests. A better understanding of how these four iterative methods act as smoothers in realistic settings, and how they are actually implemented in DD-$\alpha$AMG, is gained if we include a discussion of odd-even preconditioning, which we do in subsect.~\ref{subsubsect:oddeven_prec}.

\subsubsection{SAP}

Pseudocode for SAP is presented in alg.~\ref{alg:sap}. We explain now how SAP is realized in DD-$
\alpha$AMG. A domain decomposition of the lattice takes place to divide it in contiguous but disjoint regions consisting of four-dimensional blocks of the lattice; this domain decomposition can match the process decomposition, but in DD-$\alpha$AMG algorithmic and computational performance considerations lead to the use of a \textit{refinement} of the process decomposition when defining the domain decomposition used by SAP. Once the lattice has been decomposed into blocks, the blocks are numbered and colored (two colors, say red and black) such that no two contiguous blocks share the same color. Finally, alg.~\ref{alg:sap} runs over the red blocks first, approximately solving the linear systems defined on each block, then black blocks are processed, and these two steps are wrapped with a for loop. The block solves are done via Minimal Residual \cite{saad2003iterative}, and we write SAP($n$,$m$) to indicate $n$ outer iterations in the Schwarz smoother with $m$ iterations for the Minimal Residual solving the block-wise linear systems.

Being the goal of the smoother the rapid removal of components of the error connected to large eigenmodes of $D$, the local updates in SAP resonate more with such ``high-frequency" error to be suppressed. On the other hand, from a computational performance point of view global dot products tend to be problematic as they harm strong scaling, which is alleviated by SAP as the dot products in SAP($n$,$m$) are only of a local nature. Furthermore, having Minimal Residual running on one local SAP block at a time can possibly lead to great cache re-usage, provided the domain decomposition blocks are small enough.

\begin{algorithm}
\caption{SAP($n$,$m$) } \label{alg:sap}
\KwIn{initial guess $x^{(0)}$, matrix $D$, right hand side $b$, SAP iterations $n$, number of domain-decomposition blocks $n_b$, Minimal Residual iterations $m$}
\KwOut{$x^{(n)}$}
\For{ $i=1,...,n$}{
  $x^{(i)} = x^{(i-1)}$, $r^{(i)}=b-Dx^{(i)}$\\
  check convergence via $\|r^{(i)}\|_{2}$ and exit if done\\
  \For{$k=1,...,n_b$}{
    \If{color[k]==red}{
      perform Minimal Residual to solve $D e^{(i)}_k=r^{(i)}_k$ on a single block $k$, then $x_{k}^{(i)}=x_{k}^{(i)}+e_{k}^{(i)}$
      }
    }
    $r^{(i)}=b-Dx^{(i)}$\\
  \For{$k=1,...,n_b$}{
    \If{color[k]==black}{
      perform Minimal Residual to solve $D e^{(i)}_k=r^{(i)}_k$ on a single block $k$, then $x^{(i)}_k=x^{(i)}_k+e^{(i)}_k$
      }
    }
}
\end{algorithm}

\subsubsection{GMRES}

GMRES \cite{saad1986} is a well-known and frequently used solver for non-Hermitian systems. Via an Arnoldi process, GMRES builds a Krylov subspace $\mathcal{K}_{m}(D,r^{(0)})$ from which information is extracted to approximate the error in $D e^{(0)} = r^{(0)}$ by a residual-minimization process. For a matrix $D \in \mathbb{C}^{d \times d}$, GMRES converges in at most $d$ iterations, but in practice it does so in $n \ll d$. The value of $n$ is oftentimes prohibitively large in large-scale scientific computing simulations hence restarted GMRES($n$,$m$) is used instead, which we outline in alg.~\ref{alg:gmres}. Even with restarts, GMRES can be too expensive to use in solving linear systems; one can resort to \textit{preconditioning} \cite{saad2003iterative} in such scenarios.

A drawback of GMRES compared to SAP is the need of global dot products, leading to better strong scaling when using the latter. Furthermore, the frequent global applications of the Dirac operator in GMRES, see line 5 in alg.\ \ref{alg:gmres}, in combination with the global dot products in lines 2, 7 and 8 therein, make it difficult to have a cache-friendly execution when the local lattice, i.e.\ the lattice size per \texttt{MPI} process, is large. On the other hand, GMRES provides a faster transfer of updated data throughout the lattice, which is convenient when the system is not too ill-conditioned and the multigrid solver depends more heavily on the global convergence of the smoother.

\begin{algorithm}
\caption{GMRES($n$,$m$) } \label{alg:gmres}
\KwIn{initial guess $x^{(0)}$, matrix $D$, right hand side $b$, number of cycles $n$, cycle length $m$}
\KwOut{$x^{(n)}$}
\For{ $i = 1,{\ldots},n$}{
  $x^{(i)} = x^{(i-1)}$, $r^{(i)} = b - Dx^{(i)}$, $ \beta=h_{1,0}=\left\| r^{(i)} \right\|_2$ \\
  check convergence via $\beta$ and exit if done\\
  \For{ $k=1,...,m$}{
       $v_k = r^{(i)}/h_{k,k-1}$, $w = D v_k$\\
        \For{$j=1,$...$,k$}{
            $h_{j, k}=\left(w,v_j\right) $, $ w=w-h_{j, k} v_j$
        }
        $h_{k+1, k}=\left\|w\right\|_2$  
  }
  define $V_k=\left[v_1, \cdots, v_k\right], H_k=\left\{h_{i, j}\right\}_{1 \leq i \leq k+1,1 \leq j \leq k}$\\
  compute $y_k=\arg \min _y\left\|\beta e_1-H_k y_k\right\|_2$ and update the current iterate $x^{(i)} = x^{(i)}+V_k y_k$ \\
}
\end{algorithm}

\subsubsection{GCR} \label{subsubsect:gcr}

GCR \cite{elman1982iterative}, a Krylov method applicable as well to non-Hermitian systems, is mathematically equivalent to GMRES, minimizing also the residual in the affine space $x^{(0)} + \mathcal{K}_{m}(D,r^{(0)})$. Instead of using an orthonormal basis, the vectors $p_{k}$ from the Krylov subspace are made $D^{H}D$-orthogonal, a general condition which is constrained further in GCR by taking the next vector i.e.\ $p_{k+1}$ to be a linear combination of the current residual and the previous Krylov vectors $p_{i}, \ i \leq k$. Alg.\ \ref{alg:gcr} shows a possible formulation of GCR which restricts the number of applications of the matrix $D$ to one per iteration.

\begin{algorithm}
\caption{GCR($n,m$)} \label{alg:gcr}
\KwIn{initial guess $x^{(0)}$, matrix $D$, right hand side $b$, number of cycles $n$, cycle length $m$}
\KwOut{$x^{(n)}$}
\For{ $i = 1,{\ldots},n$}{
   $x^{(i)} = x^{(i-1)}$, $r^{(i)} = b - Dx^{(i)}$ \\
   check convergence via $\|r^{(i)}\|_{2}$ and exit if done\\
   $p_{1} = r^{(i)}$, $z_{1} = D p_{1}$ \\
   \For{ $k=1,...,m$}{
     $\delta_{k} = (z_{k},z_{k})$ \\
     $\alpha = \frac{(r^{(i)},z_{k})}{\delta_{k}}$ \\
     $x^{(i)} = x^{(i)} + \alpha p_{k}$ \\
     $r^{(i)} = r^{(i)} - \alpha z_{k}$ \\
     $w = D r^{(i)}$ \\
     \For{ $j=1,...,k$}{
       $\beta_{j} = -\frac{(w,z_{j})}{\delta_{j}}$ \\
     }
     $p_{k+1} = r^{(i)} + \sum_{j=1}^{k}\beta_{j} p_{j}$ \\
     $z_{k+1} = w + \sum_{j=1}^{k}\beta_{j} z_{j}$
   }
}
\end{algorithm}

A comparison between GMRES and GCR was undertaken when the former was introduced in \cite{saad1986}. In that work the mathematical equivalence of both methods is made clear, but furthermore the superiority of GMRES is evident due to GCR not being completely stable plus, as we can see from comparing alg.s\ \ref{alg:gmres} and \ref{alg:gcr}, the number of vectors needed in GCR is double the number for GMRES, as we need to store the $Dp_{k}$ vectors to avoid an extra application of $D$ per iteration. Moreover, a direct comparison of alg.s\ \ref{alg:gmres} and \ref{alg:gcr} in terms of work shows that GCR needs more floating point operations per iteration, another advantage in favour of GMRES.

\subsubsection{Richardson}

The last iterative method under consideration in this section is Richardson iteration \cite{richardson1911ix}. This simple method consists of approximating the correction $e^{(i)} = D^{-1}r^{(i)}$ by $e^{(i)} \approx \omega r^{(i)}$, where the weight factor $\omega$ is chosen for optimal convergence. The algorithm implementing this method is listed in alg.\ \ref{alg:richardson}. We have included the correction factor $\delta$ as a parameter that can be tuned for optimal convergence of this iterative method; we do this in this manner as the value of $\lambda_{max}$ is computed only approximately.

From alg.\ \ref{alg:richardson}, the benefits of Richardson over the other iterative methods presented in the previous three sections are clear. Its memory requirements are minimal, as besides the current iterate $x^{(i)}$, we only need an extra vector for hosting the residual $r^{(i)}$. The computation of $\lambda_{max}(D)$ can be done approximately with a few iterations of power iteration, which does not represent a large extra cost. Furthermore,
having a simple iterative method such as Richardson as a smoother opens up the possibility of undertaking a convergence analysis of the multigrid method, especially considering that, as described in subsect.\ \ref{subsect:ddalphaamg}, approximate low modes of $D$ are used in the construction of the coarse grid correction. An important drawback of Richardson compared to SAP, GMRES and GCR is that the convergence of the latter methods can be much more superior to that of Richardson in some situations. We will see, though, that when combining odd-even preconditioning (see subsect.\ \ref{subsubsect:oddeven_prec}) with these smoothers, Richardson's effectiveness as a smoother grows closer to that of the other three iterative methods considered here.

\begin{algorithm}
\caption{Richardson($n$, $\delta$) } \label{alg:richardson}
\KwIn{initial guess $x_0$, matrix $D$, right hand side $b$, number of iterations $n$, correction factor $\delta$}
\KwOut{}
approximately compute $\omega = \delta / |\lambda_{max}(D)|$\\
\For{ $i = 1,{\ldots},n$}{
   $x^{(i)} = x^{(i-1)}$, $r^{(i)} = b - Dx^{(i)} $ \\
   $x^{(i)} = x^{(i)} + \omega r^{(i)}$
}
\end{algorithm}

\subsubsection{Odd-even preconditioning} \label{subsubsect:oddeven_prec}

To close with this subsection, we describe now odd-even preconditioning. This is an important technique always present in LQCD solvers, and in particular is integrated with the smoothers in DD-$\alpha$AMG.

Let us assume that we color the four-dimensional LQCD grid with two colors, say red and black. Due to $D$ having not further than nearest-neighbor interactions, the two regions in this domain decomposition are decoupled. An alternative way to do this is to label each lattice site by its four-dimensional coordinates as $(x_{1},x_{2},x_{3},x_{4})$, which allows us to do an odd-even decomposition, equivalent to the red-black one, by splitting the lattice in two disjoint regions consisting of \textit{even} and \textit{odd} sites, where a site is even if $x_{1}+x_{2}+x_{3}+x_{4}$ is even and analogously for odd sites.

If one reorders the vectors in the linear system $Dx=b$ by even sites first and odd ones second, then the linear system takes the form
\begin{equation}
    \begin{pmatrix}
    D_{ee} & D_{eo}\\
    D_{oe} & D_{oo}
    \end{pmatrix}
    \begin{pmatrix}
    x_{e}\\
    x_{o}
    \end{pmatrix}
    =
    \begin{pmatrix}
    b_{e}\\
    b_{o}
    \end{pmatrix}.
\end{equation}

With these rearrangements, we can then write the inverse of $D$ conveniently as \cite{saad2003iterative}
\begin{equation} \label{eq:Dinv_in_factorized_form}
    D^{-1}=
    \begin{pmatrix}
    I & 0\\
    -D_{oo}^{-1}D_{oe} & I
    \end{pmatrix}
    \begin{pmatrix}
    D_{S}^{-1} & 0\\
    0 & D_{oo}^{-1}
    \end{pmatrix}
    \begin{pmatrix}
    I & -D_{eo}D_{oo}^{-1}\\
    0 & I
    \end{pmatrix},
\end{equation}
with the Schur complement
\begin{equation} \label{eq:schur_complement}
    D_{S} = D_{ee} - D_{eo} D_{oo}^{-1} D_{oe}.
\end{equation}

In the particular case of LQCD, $D_{ee}$ and $D_{oo}$ are block-diagonal matrices with $6 \times 6$ blocks, hence explicitly storing the data for $D_{oo}^{-1}$ is cheap.

\begin{figure*}
\includegraphics[width=0.325\textwidth]{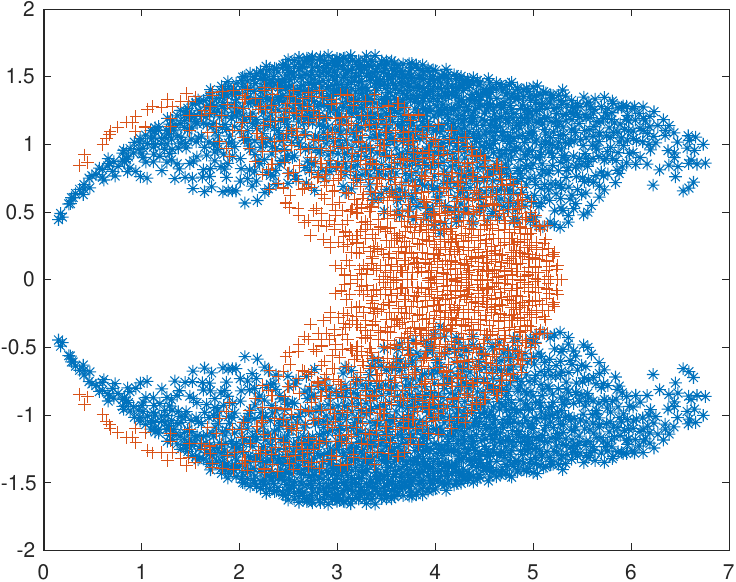}\includegraphics[width=0.335\textwidth]{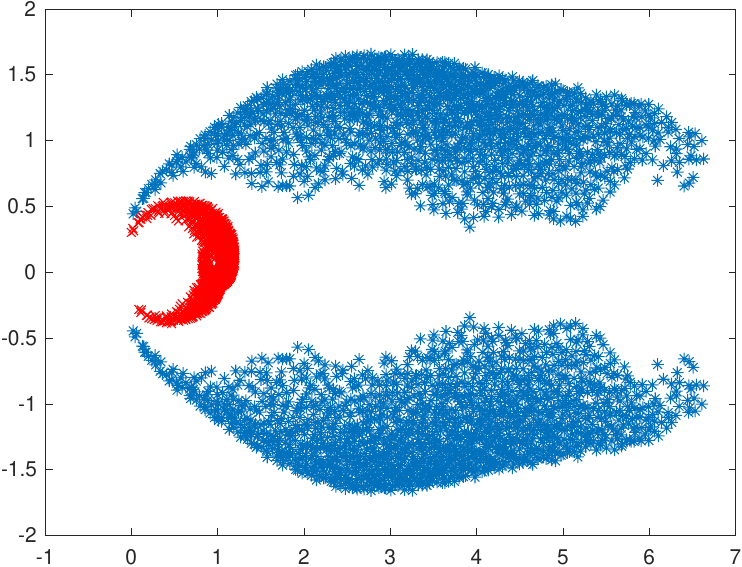}\includegraphics[width=0.335\textwidth]{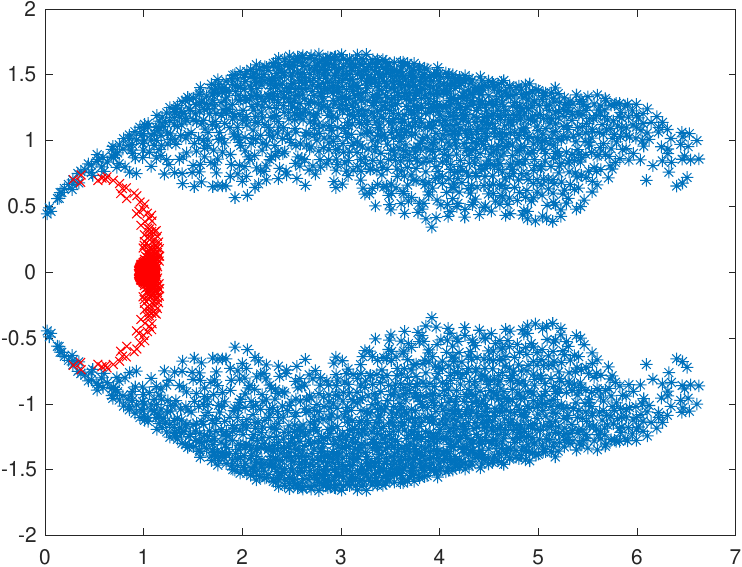}
\caption{$4^4$ lattice. Spectra. \textit{Left}: spectrum of full $D$ and spectrum of $D_{S}$ (Schur complement). \textit{Center}: spectrum of full $D$ and $D p_{R,3}(D)$ with $p_{R,3}(D)$ the Richardson polynomial of degree three. \textit{Right}: spectrum of full $D$ and spectrum of $D M_{3}$ where $M_{3}$ is the factorization of $D^{-1}$ via Schur complement with the inverse of the Schur complement $D_{S}^{-1}$ replaced by $p_{R,3}(D_{S})$.} \label{fig:oddeven_spectra_with_Richardson}
\end{figure*}

We can illustrate the usefulness of odd-even preconditioning by taking a look at spectra of both $D$ and $D_{S}$. This is done in the left panel in fig.\ \ref{fig:oddeven_spectra_with_Richardson}, where we have used a Dirac matrix corresponding to a $4^4$ lattice. In there, the blue spectrum, i.e.\ the one extending beyond 6.0, is $\mathrm{spec}(D)$, and the other one is $\mathrm{spec}(D_{S})$. We see then that $D_{S}$ is not only less ill-conditioned than $D$, but the part of the spectrum with largest eigenvalues in magnitude has moved closer to zero and it has become more uniformly distributed, and the smallest eigenvalues in magnitude have moved further away from zero and become more scattered, becoming then an easier linear system for iterative methods, such as the ones presented in previous sections, to deal with.

There are two ways in which odd-even preconditionining is combined with the smoother in DD-$\alpha$AMG. The first is the same for GMRES, GCR and Richardson, while the second one is particular to SAP.

When combining odd-even preconditioning with GMRES, GCR or Richardson, the whole lattice is reordered in an odd-even manner, and then the chosen iterative method is applied to the solves involving the Schur complement. On the other hand, when used in conjunction with SAP, it is not the whole lattice what is reordered but rather each of the domain decomposition blocks, see alg.\ \ref{alg:sap}, and then minimal residual is used for solving the linear systems with the Schur complement of each block. More specifically, odd-even preconditioning is applied to the block linear systems in lines 5 and 9 in alg.\ \ref{alg:sap}.

In the case of SAP, it is clear that odd-even preconditioning does not change the smoothing nature of it, as it only helps the block solves in SAP to converge faster, but a question at this point is whether it helps GMRES, GCR and Richardson in being better smoothers. We answer this question via the spectra displayed in the center and right panels in fig.\ \ref{fig:oddeven_spectra_with_Richardson}. To this purpose, let us first state that one can write the Richardson polynomial, i.e.\ the polynomial in $x^{(k)} = x^{0} + p_{R,k}(D) r^{(0)}$ after $k$ iterations of Richardson, as $p_{R,k}(D) = ( I-(I-\omega D)^{k} )D^{-1}$. Then, in the center panel we show in red $\mathrm{spec}\left(Dp_{R,3}(D)\right)$, i.e.\ the shrunk spectrum. Finally, in the right panel we plot $\mathrm{spec}\left(DM_{3}\right)$, where $M_{3}$ is $D^{-1}$ in the factorized form in eq.\ \ref{eq:Dinv_in_factorized_form} with $D_{S}^{-1}$ replaced by $p_{R,3}(D_{S})$. It is then clearly beneficial to have a smoother with odd-even preconditioning on the whole lattice and an iterative method such as GMRES, GCR or Richardson acting on the inner linear system with the Schur complement, as the spectrum in the right panel shows a better mapping to 1.0, as well as a distancing of the spectrum from zero and a scattering of the low modes.

\subsection{\texttt{CUDA} and \texttt{HIP}}

\texttt{CUDA} \cite{NVIDIACU33:online} (Compute Unified Device Architecture) and \texttt{HIP} \cite{HIPdocum88:online,HIP_GitHub} (Heterogeneous-Compute Interface for Portability) are parallel computing platforms that allow developers to harness the power of GPUs to accelerate their applications. Both \texttt{CUDA} and \texttt{HIP} provide the Single-Instruction-Multiple-Thread (SIMT) parallel programming model, and offer  similar mechanisms for managing memory on the GPU, including allocation, data movement between CPU and GPU, and synchronization. They also have rich basic libraries, such as linear algebra, image processing, and deep learning, as well as extensive developer communities and ecosystems.

\texttt{CUDA} is developed by NVIDIA and specific to NVIDIA GPUs, which uses NVIDIA's proprietary \texttt{CUDA} C/C++ programming language, while \texttt{HIP}  is an open-source alternative developed by AMD. \texttt{HIP} aims for portability across different GPU architectures, including both AMD and NVIDIA GPUs, which supports C/C++, making it easier to port \texttt{CUDA} applications to \texttt{HIP}.

\texttt{CUDA} has been around since 2006, giving it a significant head start. It has a vast ecosystem with mature libraries, tools, and extensive community support. However \texttt{CUDA} ties applications tightly to NVIDIA GPUs, limiting their portability across different GPU architectures. Therefore it is understandable that the NVIDIA GPUs optimized for \texttt{CUDA} usually provide better performance for certain workloads, particularly in areas like deep learning and scientific computing.

One of the most significant advantages of \texttt{HIP} is its goal of cross-platform portability. It enables developers to write code that can be compiled and executed on different GPU architectures, allowing applications to be easily transferred between AMD and NVIDIA platforms. Encouraged by the open-source strategy, HIP benefits from community contributions, faster bug fixes, and wider adoption by other GPU vendors. Compared to the vendor lock-in strategy of \texttt{CUDA}, \texttt{HIP}'s support for NVIDIA GPUs is not as comprehensive as \texttt{CUDA}'s.

Overall, \texttt{CUDA} is preferred when targeting NVIDIA GPUs with a mature ecosystem, while \texttt{HIP} offers more flexibility and portability across different GPU architectures. The choice depends on the specific requirements, target hardware, and development preferences of the project. In this work, we take code developed in \texttt{CUDA} and translate it to be compiled and run with \texttt{HIP} on the ORISE supercomputer.

\section{Numerical Experiments}\label{sect:num_exps}

We start our numerical experiments by comparing the four iterative methods described in sect.~\ref{sect:amg_and_orise} as smoothers in DD-$\alpha$AMG, and their interplay with odd-even preconditioning. We do this first on CPUs only in subsect.\ \ref{subsect:smoothers_comp_cpu_only}. The DD-$\alpha$AMG code that we have ported with \texttt{CUDA} and translated into \texttt{HIP} is then run on ORISE, and the results of this are presented in subsect.~\ref{subsect:ddalphaamg_on_orise_results}.
Finally, we present results in subsect.~\ref{subsect:avx2_on_coarser_levels_results} on the acceleration of some coarser-level kernels via AVX vectorization. When reporting timings for solving linear systems, we report the solve-phase time only; giving timings for the setup phase of the multigrid solver too would imply a full discussion of the setup phase, which is beyond the scope of this paper.

The details of the configuration that we have used throughout most of our numerical experiments are listed in tab.~\ref{tab:config}. It is important at this point to notice the need of multigrid for this configuration. Its pion mass is 135 MeV, corresponding to its physical value, which will lead to quite ill-conditioned linear systems. To exemplify this, we have solved a linear system for this configuration with DD-$\alpha$AMG both with BiCGStab and three-level multigrid as preconditioners of FGMRES \cite{rottmann2016adaptive}, with strange and light quark masses. The parameters for the two-level multigrid method can be found in tab.\ \ref{tab:parameter}. This run was done with CPU-only code on 9 nodes of ORISE with a relative residual tolerance of $10^{-10}$. For strange quark mass, using BiCGStab takes 23.9 seconds while with multigrid it takes 11.3 seconds. When we switch to light quark mass the times become 590.2 seconds and 31.5 seconds, respectively. These times illustrate how simulations might become unfeasible without multigrid, as oftentimes the overall execution time is dominated by the solution of the linear systems studied here.

\subsection{The smoothers: CPU-only comparisons}\label{subsect:smoothers_comp_cpu_only}

We have extended the capabilities of DD-$\alpha$AMG first by implementing CPU versions of GCR and Richardson, both with and without odd-even preconditioning. Before this work, DD-$\alpha$AMG was capable of running only with SAP or GMRES as the smoother. We compare now these four smoothers to assess their effectiveness and differences in practice. To do this comparison, we have run a two-level multigrid method with all of these smoothers under similar conditions such as initial random seed, parallelization, restart length of Arnoldi processes, etc. The results of these runs can be found in tab.\ \ref{tab:smoothers_cpu_only_comparison}, and the parameters of the two-level method in tab.\ \ref{tab:parameter}.

In order to compare pros and cons of the different smoothers, we have tested the four combinations of with or without odd-even preconditioning and strange or light quark mass. We can see from tab.\ \ref{tab:smoothers_cpu_only_comparison} that, without odd-even preconditioning, SAP is the most effective smoother both in terms of convergence and execution time of the multigrid solver, partly due to dealing most effectively with the suppression of the high-frequency components of the error. Furthermore, as we only do local solves in SAP, caching is quite favorable compared to GMRES and GCR, leading to a lower time per iteration in the former iterative method. When comparing GMRES to GCR, the latter takes longer per iteration, which matches the discussion in \cite{elman1982iterative} and subsect.\ \ref{subsubsect:gcr}. Overall, in this scenario of no odd-even preconditioning, Richardson is not competitive enough as it leads to very low convergence of the whole multigrid method in the strange quark mass case, and in the light quark mass regime to no convergence at all.

Things change considerably for Richardson, compared to the other three smoothers, when we switch odd-even preconditioning on. In this case Richardson leads to multigrid solves almost as fast as SAP and GMRES do, for both values of quark mass; the difference is only of 10\% with respect to SAP. Two clear advantages of odd-even Richardson over the other three smoothers is that, besides its competitive effectiveness as a smoother in terms of iteration count and execution time of the multigrid solver, its memory requirements are minimal and it is the simplest of these methods to implement by far, which are very attractive features in codes that try to be portable and simple but fast and low memory-consuming at the same time.

We note that, although we have used different values of Richardson's $\delta$ in some of the different cases in tab.\ \ref{tab:smoothers_cpu_only_comparison}, a value of $\delta=1.5$ seems general enough in terms of giving approximate optimal convergence.

We now explore these smoothers on GPUs. Due to the larger time and memory required by GCR compared to the other three smoothers considered above, we exclude this method from our discussions in the next section.

\begin{table*}[h]
    \centering
    \caption{Comparative timings of different smoothers in DD-$\alpha$AMG, namely SAP, GMRES, GCR and Richardson iteration, for both strange and light quark masses, with CPU-only code. A two-level method has been used in all cases. All runs were done on 4 nodes in ORISE. The times are in seconds. Via the dashes we indicate that the method did not converge.}
   \label{tab:smoothers_cpu_only_comparison}
   \begin{tabular}{ccccccc}
     \hline
      & mass                     & smoother       & FGMRES iters & smoothing time & coarse time & solve time  \\ \hline
      &                          & SAP(3,4)       & 22           & 169.83         & 34.33       & 245.27      \\
      & \multirow{2}{*}{$m_{s}$} & GMRES(3,4)     & 24           & 278.12         & 37.88       & 364.21      \\
      &                          & GCR(3,4)       & 24           & 354.89         & 36.75       & 437.86      \\
      &                          & Richardson(12,1.7) & 95          & 750.03         & 92.08        & 1028.93      \\
     non odd-even &&&&&&\\[-1.5ex]
     \cline{2-7}
     &&&&&&\\[-1.5ex]
      &                          & SAP(3,4)       & 43   & 337.35 & 693.92 & 1122.45 \\
      & \multirow{2}{*}{$m_{l}$} & GMRES(3,4)     & 55   & 639.47 & 939.22 & 1693.01 \\
      &                          & GCR(3,4)       & 53   & 797.05 & 893.82 & 1812.04 \\
      &                          & Richardson(12,1.5) & -    & -      & -      & -       \\
          &&&&&&\\[-1.5ex]
          \hline
          &&&&&&\\[-1.5ex]
          &                           & SAP(3,4)       & 16  & 70.3   & 6.73  & 96.94  \\
          &  \multirow{2}{*}{$m_{s}$} & GMRES(3,4)     & 13  & 73.56  & 6.23  & 96.70  \\
          &                           & GCR(3,4)       & 13  & 92.15  & 6.08  & 115.51 \\          
          &         & Richardson(12,1.5) & 15   & 72.02  & 7.45 & 100.46      \\  
     odd-even     &&&&&&\\[-1.5ex]
     \cline{2-7}
          &&&&&&\\[-1.5ex]
          &         & SAP(3,4)        & 23  & 100.58  & 166.72  & 300.26  \\
          & \multirow{2}{*}{$m_{l}$}         & GMRES(3,4)      & 23  & 131.63  & 145.27  & 310.19  \\
          &    
          & GCR(3,4)         & 23  & 171.27   & 153.03   & 366.15   \\          
          &         & Richardson(12,1.5) & 28   & 135.05  & 163.21 & 335.84      \\  
          \hline
   \end{tabular}
   \end{table*}

\subsection{DD-$\alpha$AMG via \texttt{HIP} on ORISE}\label{subsect:ddalphaamg_on_orise_results}

Previous to this work, the only functions in DD-$\alpha$AMG ported via \texttt{CUDA} were finest-level SAP and the finest-level Dirac operator in double precision. We have extended this by porting the odd-even preconditioned versions of GMRES and Richardson, also via \texttt{CUDA}. To accomplish this we have made use of some of the existing \texttt{CUDA} kernels, previously developed when porting the finest-level SAP and double precision Dirac operator, and created some kernels of our own, in order to first have a Schur complement (see eq.\ \ref{eq:schur_complement}) capable of running on GPUs. With a GPU version of the Schur complement as base, both odd-even GMRES and Richardson were built around it, with GMRES clearly needing more development time due to its inner products, least squares minimization, etc.

Again, as in the CPU case, we have compared SAP, GMRES and Richardson under similar conditions, the results of which are presented in tab.\ \ref{tab:comp-cpu-gpu}.

Due to having already thoroughly compared the different smoothers within the multigrid context in tab.\ \ref{tab:smoothers_cpu_only_comparison}, we focus now in tab.\ \ref{tab:comp-cpu-gpu} on the time taken by the smoother only; these times come from averaging smoothing time over multiple multigrid solves. We display the best CPU time versus the best GPU time in tab.\ \ref{tab:comp-cpu-gpu}. The superiority of SAP is clear not only on CPUs but also on GPUs, in the former case due to great CPU cache usage, and in the latter due to the independence of the SAP blocks leading to the possibility of launching operations from all those independent blocks asynchronously on GPUs. A further advantage of SAP over GMRES is that dot products are of a local rather than a global nature on the lattice.

Although odd-even Richardson consists mostly of the application of a global Schur complement, tab.\ \ref{tab:comp-cpu-gpu} shows that the average time for one call of this iterative method approximately matches that of SAP on GPUs. This keeps Richardson again competitive when considered as a smoother for multigrid on GPUs, and this would be, again, the smoother of choice when simplicity of implementation and low memory-consumption are of utmost importance.

\begin{table}
    \centering
    \caption{Comparison of the different smoothers on GPU to CPU. Running on 4 nodes, 4 MPI processes per node, and 1 GPU and 4 OpenMP threads per process. The times are in seconds and correspond to the average time for one smoother call.}
    \label{tab:comp-cpu-gpu}
    \begin{tabular}{cccc}
    \hline
    smoother   & CPU     & GPU     & speedup   \\ \hline
    SAP(3,4)   & 4.39    & 0.33    & 13.30      \\
    GMRES(3,4) & 5.66    & 0.45    & 12.60     \\
    Richardson & 4.80    & 0.36    & 13.33     \\ \hline
    \end{tabular}
\end{table}

\subsection{CPU improvements}\label{subsect:avx2_on_coarser_levels_results}

With the finest-level smoother and finest-level Dirac operator running on GPUs in the hybrid CPU+GPU implementation of DD-$\alpha$AMG \cite{wuppertalGPUDDalphaAMG}, a possible further improvement of the solver comes from looking at coarse-grid computations. Since its original version \cite{wuppertalDDalphaAMG}, vectorization via \texttt{SSE} is used in DD-$\alpha$AMG to speed up the execution of the most demanding operations, e.g.\ in the application of the Dirac operator at different levels. We have extended some of the vectorized coarse-grid kernels to support \texttt{AVX2} and \texttt{AVX-512} as well.

At the $l$-th level, for $l>1$ the interaction between two neighboring sites is encoded as a $2N_{v} \times 2N_{v}$ dense complex matrix, where $N_{v}$ is the number of test vectors chosen at level $l-1$. The structure of the data and the operations implied in these small and dense matrix-vector products have a natural compatibility with the use of vectorization schemes such as \texttt{SSE}. Being the cache line size typically 64 B, the corresponding 16 numbers in single precision are a natural match with the registers in \texttt{SSE}, \texttt{AVX2} and \texttt{AVX-512}, where the sizes of the vectorization registers are 4, 8 and 16, respectively, and furthermore the value of $2N_{v}$ is typically a multiple of these vectorization register sizes.

Although all these data sizes align very well, when applying the Dirac operator $D_{l}$, $l>1$, one goes over a sequence of these small and dense matrix-vector products without reusing data from one matrix-vector multiplication to the next. Hence, in a single application of the Dirac operator, the execution time is highly bound by memory bandwidth. In such a case of applying $D_{l}$ only once, upgrading from \texttt{SSE} to \texttt{AVX2} or \texttt{AVX-512} will probably be of very little benefit.
But if $D_{l}$ is applied many times, whether we benefit from \texttt{AVX} or not will depend on the data needed by $D_{l}$ and how this compares to the cache size.

If we take $N_{v}=24$, one of those small dense matrices within $D_{l}$ ($l>1$) takes up $48 \times 48 \times 2 \times 4 = 18.0$ KB of data in single precision. An Intel(R) Xeon(R) Platinum 8180 node, for example, has 32 KB of L1 cache per core, which allows hosting only one of those small dense matrices, but its L3 cache, on the other hand, is 38.5 MB, which fits a maximum of 2,190 such matrices. The Dirac operator requires the storage of roughly nine of those $2N_{v} \times 2N_{v}$ blocks per block-row. With each block-row mapping directly to a single lattice site, the maximum number of sites that we can place in L3 cache is then roughly 243. When running an iterative method such as GMRES, see alg.\ \ref{alg:gmres}, the Dirac operator is applied once per iteration, therefore if the iterative method takes many iterations and all the data required by this method fits in cache, we will see benefits by moving up from \texttt{SSE} to \texttt{AVX}. The data required by a method such as GMRES will be that of the Dirac operator plus multiple vectors, e.g.\ the Arnoldi vectors and some buffers. We verify now these considerations by means of DD-$\alpha$AMG and the vectorization upgrades that we have included in it via \texttt{AVX}.

On the same Xeon node as before, we run a two-level method with DD-$\alpha$AMG for a gauge configuration with an underlying lattice of size $8^4$. We do this for two aggregation sizes, $2^4$ and $4^4$, leading to coarse grids of $4^4$ and $2^4$, respectively. A coarse grid of $4^4$ has a total of 256 lattice sites which, as discussed above, does not fit in L3 cache. On the other hand, having a coarse grid of $2^4$ a total of 16 sites, the whole data needed by its Dirac operator fits in L3. We have built a test where we call the coarse-grid Dirac operator a total of 10,000 times. We have run this experiment multiple times, averaged over the resulting execution times, and displayed the results in tab.s\ \ref{tab:avx_vs_sse_coarse_grid_Dirac_4to4} and \ref{tab:avx_vs_sse_coarse_grid_Dirac_2to4}. The overhead in communications coming from the halo exchanges in the application of the Dirac operator does not allow us to see the bare impact of our vectorization upgrades. Hence, we have also run tests with a modified Dirac operator, where we have disabled those nearest-neighbor data transfers, the timings of which we also include in tab.s \ref{tab:avx_vs_sse_coarse_grid_Dirac_4to4} and \ref{tab:avx_vs_sse_coarse_grid_Dirac_2to4} for coarse grids of $4^4$ and $2^4$, respectively.

\begin{table}
    \centering
    \caption{Times and speedups of \texttt{AVX2} and \texttt{AVX-512} compared to \texttt{SSE} for a coarse-grid Dirac operator corresponding to a $4^4$ lattice. All the times are in seconds and correspond to an average of the time that it takes to apply the coarse-grid Dirac operator in single precision a total of 10,000 times. The numbers in parenthesis for \texttt{AVX2} and \texttt{AVX-512} are speedups over the \texttt{SSE} case. The first column indicates whether we have turned off halo exchanges in the Dirac operator or not. }
    \label{tab:avx_vs_sse_coarse_grid_Dirac_4to4}
    \begin{tabular}{ccccc}
    \hline
    comm.s                 & \# MPI   & SSE   & AVX2        & AVX-512     \\
                           & Proc.s   &       &             &             \\ \hline
    \multirow{5}{*}{ON}    & 2        & 13.9  & 9.7  (1.43) & 9.5 (1.46)  \\
                           & 4        & 7.64  & 5.39 (1.42) & 5.3 (1.44)  \\
                           & 8        & 4.5   & 3.47 (1.3)  & 3.2 (1.41)  \\
                           & 16       & 2.4   & 2.01 (1.19) & 1.89 (1.27) \\
                           & 32       & 2.33  & 1.96 (1.19) & 1.91 (1.22) \\ \hline
    \multirow{5}{*}{OFF}   & 2        & 12.9  & 9.2 (1.4)   & 9.02 (1.43) \\
                           & 4        & 7.0   & 4.9 (1.43)  & 4.64 (1.5)  \\
                           & 8        & 3.7   & 2.4 (1.54)  & 2.37 (1.56) \\
                           & 16       & 2.1   & 1.6 (1.3)   & 1.53 (1.37) \\
                           & 32       & 1.3   & 1.03 (1.26) & 0.97 (1.34)            \\ \hline
    \end{tabular}
\end{table}

\begin{table}
    \centering
    \caption{Times and speedups of \texttt{AVX2} and \texttt{AVX-512} compared to \texttt{SSE} for a coarse-grid Dirac operator corresponding to a $2^4$ lattice. All the times are in seconds and correspond to an average of the time that it takes to apply the coarse-grid Dirac operator in single precision a total of 10,000 times. The numbers in parenthesis for \texttt{AVX2} and \texttt{AVX-512} are speedups over the \texttt{SSE} case. The first column indicates whether we have turned off halo exchanges in the Dirac operator or not.}
    \label{tab:avx_vs_sse_coarse_grid_Dirac_2to4}
    \begin{tabular}{ccccc}
    \hline
    comm.s                 & \# MPI   & SSE   & AVX2        & AVX-512     \\
                           & Proc.s   &       &             &             \\ \hline
                           & 2        & 0.97  & 0.67 (1.45) & 0.62 (1.56) \\
    \multirow{2}{*}{ON}    & 4        & 0.62  & 0.42 (1.48) & 0.31 (2.0)  \\
                           & 8        & 0.4   & 0.28 (1.43) & 0.24 (1.67) \\ \hline
                           & 2        & 0.86  & 0.56 (1.54) & 0.48 (1.8)  \\
    \multirow{2}{*}{OFF}   & 4        & 0.47  & 0.26 (1.8)  & 0.19 (2.47) \\
                           & 8        & 0.26  & 0.16 (1.6)  & 0.11 (2.36) \\ \hline
    \end{tabular}
\end{table}

The time reduction when going from \texttt{AVX} to \texttt{SSE} is larger in tab.\ \ref{tab:avx_vs_sse_coarse_grid_Dirac_2to4} than in tab.\ \ref{tab:avx_vs_sse_coarse_grid_Dirac_4to4}, as expected. This is simply due to having more cache misses in the latter case, where the whole data required by $D_{l}$ cannot be fully stored in L3. It is important to note that \texttt{AVX-512} usually runs with a reduced frequency compared to \texttt{AVX2}, hence the speedup attained with the former is far from being double than with the latter, matching our results in in tab.\ \ref{tab:avx_vs_sse_coarse_grid_Dirac_2to4}.

Finally, when we run the whole DD-$\alpha$AMG solver for this same $8^4$ problem with \texttt{SSE}, an aggregation of $4^4$ and four MPI processes, the coarse grid takes up a total time of 0.0825 seconds, while with \texttt{AVX2} and \texttt{AVX-512} it takes 0.0616 and 0.0517, respectively. Multiple overheads and regions not upgraded by \texttt{AVX} in GMRES are the reason behind not seeing speedups of 1.48 and 2.0 as suggested by tab.\ \ref{tab:avx_vs_sse_coarse_grid_Dirac_2to4}. This motivates the use of polynomial preconditioning in GMRES \cite{loe2022toward} for the Dirac operator, as has been implemented and analyzed in \cite{espinoza2023coarsest}. Moreover, the superior speedups obtained in tab.s \ref{tab:avx_vs_sse_coarse_grid_Dirac_4to4} and \ref{tab:avx_vs_sse_coarse_grid_Dirac_2to4} when switching communications off suggests that using communication avoiding Krylov methods \cite{hoemmen2010communication} would be of benefit.

We note from tab.s \ref{tab:avx_vs_sse_coarse_grid_Dirac_4to4} and \ref{tab:avx_vs_sse_coarse_grid_Dirac_2to4} the non-uniform speedups due to vectorization as we change the parallelism in the different runs. One way to possibly obtain better and more uniform (i.e.\ parallelism-independent) vectorization performance is by means of extending the solver to support multiple right hand sides \cite{yamamoto2022implementation}; when applying the Dirac operator on multiple vectors, one has to apply each of the $2N_{v} \times 2N_{v}$ blocks on multiple vectors, which can be made a highly cache-friendly operation.

Finally, we note that the use of \texttt{AVX} was of small benefit on ORISE, where our numerical tests showed speedups of 1.1 at best in comparison with the \texttt{SSE} implementation. We credit this to the particularly large time taken up by fetching data from main memory on that machine.

\section{Conclusions and future work} \label{sect:conclusions_and_future_work}

We have extended the CPU-only version of the DD-$\alpha$AMG solver to allow for two new smoothers, namely GCR and Richardson. In doing so we have shown that, due to time and memory reasons, GCR is not a good option. On the other hand, Richardson emerged as a very attractive alternative due to its simplicity of implementation, low memory requirements and competitive effectiveness. Furthermore, from an existing CPU+GPU version of DD-$\alpha$AMG with SAP ported via \texttt{CUDA}, we have ported GMRES and Richardson, which strengthens the superiority of SAP as a smoother not only on CPUs but also on GPUs, and at the same time reaffirms the potential and possible superiority, in some ways, of Richardson as a smoother for multigrid on both CPUs and heterogeneous architectures. Finally, coarse-grid expensive CPU kernels have been made faster via \texttt{AVX2} and \texttt{AVX-512}. A thorough analysis has been included here on the benefits of these improvements for DD-$\alpha$AMG.

Upcoming work derives from the improvements and extensions outlined above. For example, the use of Richardson as a smoother in other LQCD multigrid solver is one such path. Another possible direction of work is the use of lower precision, e.g.\ half, on coarser levels in DD-$\alpha$AMG in combination with advanced vectorization such as \texttt{AVX-512}.

\section{Acknowledgments}
We would like to thank Yibo Yang for providing us with the gauge configuration. Gustavo Ramirez-Hidalgo's work is supported by the German Research Foundation (DFG) research unit FOR5269 “Future methods for studying confined gluons in QCD”.

\appendix

\section{Configuration and solver parameters}

The gauge configuration that we have used is listed in tab.~\ref{tab:config}. The parameters used in DD-$\alpha$AMG are listed in tab.~\ref{tab:parameter}.

\begin{table}[H]
        \centering
        \caption{Configuration used together with its parameters. Provided by C11P14L\cite{LIU2023137941}.}\label{tab:config}
        \begin{tabular}{cc}
        \hline
        Lattice size ($N_t \times N_s^3$)    & $96\times 48^3$  \\
        Pion mass ($m_{\pi}$)                & 135              \\
        Clover term ($c_{sw}$)               & 1.16058719       \\
        light mass ($m_{l}$)                 & -0.2825          \\
        strange mass ($m_{s}$)               & -0.2310          \\ \hline
        \end{tabular}
\end{table}

\begin{table}[H]
    \centering
    \caption{Default parameters for two-level DD-$\alpha$AMG with SAP as smoother. These are the values used when solving with light quark mass $m_{l}$. The value of \texttt{number of setup iterations} changes to 3 for the strange quark mass.}
    \label{tab:parameter}
    \begin{tabular}{ccc}
    \hline
    parameter                          & $l=1$      & $l=2$      \\\hline
    number of setup iterations         &   5        & -          \\
    number of test vectors             &   20       & -          \\
    aggregate size                     & $4^4$      & -          \\
    restart length of (F)GMRES         & 10         & 200         \\
    maximum restarts of (F)GMRES       & 10         & 10         \\
    (F)GMRES rel.\ residual tolerance  & $10^{-10}$ & $10^{-1}$  \\
    post-smoothing steps               & 3          & -          \\
    Minimal Residual iterations        & 4          & -          \\\hline
    \end{tabular}
\end{table}

\bibliographystyle{model1-num-names}

\bibliography{cas-refs}

\end{document}